\documentclass[12pt]{amsart}
\usepackage{amsfonts,amsthm}
\usepackage{amssymb,amsmath}
\usepackage{amsfonts,mathrsfs}
\usepackage{xcolor}

\begin{document}

\noindent{\bf Felix Klein's ``Zu Hilberts erster Note \"uber die Grundlagen der Physik'': an English translation}

\bigskip
\centerline{Chiang-Mei Chen$^{1,2}$, James M. Nester$^{1,3,4}$ and Walter Vogel$^5$}

\medskip

\noindent $^1$ Department of Physics, National Central University, Chungli 32001, Taiwan

\noindent $^2$ Center for High Energy and High Field Physics (CHiP), National Central University,
Chungli 32001, Taiwan

\noindent $^3$ Graduate Institute of Astronomy, National Central University, Chungli 32001, Taiwan

\noindent $^4$ Leung Center for Cosmology and Particle Astrophysics, National Taiwan University, Taipei 10617, Taiwan

\noindent $^5$ Department of Chemistry, National Central University, Chungli, 32001, Taiwan,

\medskip
\noindent email: cmchen@phy.ncu.edu.tw, nester@phy.ncu.edu.tw, vogelw@ncu.edu.tw

\bigskip

\noindent{\textbf{Abstract}}

\emph{We present an English translation of a 1918 paper by Felix Klein.}

\section{Translator's Preface}
In 1918 Felix Klein published a work~\cite{Klein} about a famous paper by David Hilbert: The Foundations of Physics~I~\cite{Hilbert}. Klein presented his work in the form of a dialogue with Hilbert based on their correspondence. Mainly it concerned the energy of the gravitational field, which had been---and, notwithstanding considerable progress, still remains a century later---an unsettled issue.  It includes Klein's simplified derivation of Hilbert's energy vector, which he, Einstein and others were trying to understand, and some remarks regarding Einstein's gravitational energy momentum density. It was a statement by Hilbert that appears in this dialogue: ``I maintain that for \textit{general} relativity, i.e. in the case of the \textit{general} invariance of the Hamiltonian function, energy equations which in your sense correspond to the energy equations of the orthogonally invariant theories do not exist at all; indeed, I would even call this fact a characteristic feature of the general theory of relativity. For my assertion the mathematical proof should be possible'', which led to the investigation of Emmy Noether~\cite{Noether} containing her two celebrated theorems regarding symmetry in dynamical systems.  For the detailed story concerning the exchanges of ideas between Einstein, Hilbert and Klein (especially concerning gravitational energy) and the inception of Noether's theorems see Refs.~\cite{Brading05, KS, Rowe99}.

Some years ago the senior member of our team (JMN) began, relying on his long unused undergraduate German and Google Translate, to make a translation of this paper of Klein and a follow up work of his on gravitational energy.    We are fortunate to be able to have recently acquired the help of a native German speaker (WV) to refine our efforts into a presentable form. We feel that our translation has now finally reached a form where it can be useful to others, and so want to share it with anyone who may be interested.

The page by page layout, the equation numbers, and footnotes in this version of our translation are from the paper as it appears in Vol.~1 of Klein's collected works~\cite{Klein}---so anyone who cares to can easily compare our translation with the original. We chose to follow the Klein collected works version as it includes some additional footnotes that do not appear in the journal version (readers may find those regarding Emmy Noether especially interesting). Our translation is a work in progress. We welcome corrections and comments on the translation and on any errors.

\newpage

{\textbf{XXXI. On Hilbert's First Note on the Foundations of Physics.}%
\footnote{ G\"ottinger Nachrichten, Math.-phys. class, (1915), p.~395--407 (Communication of November 20, 1915).}
}\\

\footnotesize{[News of the Kgl. Society of Sciences at G\"ottingen. Mathematical-physical class, (1917). Submitted at the meeting of 25 January 1918.]}

\section[I]{From a letter from F. Klein to D. Hilbert.}

\dots By carefully studying your note, I have noticed that the intermediate calculations which you made can be considerably shortened by the use of the ordinary Lagrange variation theorem, and in this way one can get a more accurate insight into the importance of the conservation law which you set up for your energy vector. In the following discussion of my considerations, I shall continue, as far as possible, with your terms of reference, except that, for the sake of clarity, I distinguish the world parameters $w$ by \textit{upper} indices:
$$w^I, w^{II}, w^{III}, w^{IV}$$
and indeterminate indices by Greek letters. This makes it much easier to compare with the parallel developments of Einstein, about which I also have to make a few remarks.

1. I immediately begin, following your note on page 404, with the introduction of the two integrals which I call $I_1$ and $I_2$:
\begin{equation}%1
I_1 = \int K d\omega, \quad I_2 = \int L d\omega,
\end{equation}
where $d\omega$ is the invariant space element
$$ d\omega = \sqrt{g} \cdot dw^I \dots dw^{IV}. $$
Here, $K$ is the fundamental local invariant of the underlying $ds^2$, which is written using Riemann's four-index symbols as follows:
\begin{equation}%2
K = \sum_{\mu, \nu, \varrho, \sigma} (\mu \nu, \varrho \sigma) (g^{\mu \varrho} g^{\nu \sigma} - g^{\mu \sigma} g^{\nu \varrho}),
\end{equation}
\newpage

554 \hfil {On the Erlangen program.}\hfil

\bigskip

\noindent but for $L$, as I do not focus on the generality of physical requirements, I will write the simplest value according to p.~407 of your work:
\begin{equation}%3
L = \alpha Q = - \alpha \sum_{\mu, \nu, \varrho, \sigma} (q_{\mu\nu} - q_{\nu\mu}) (q_{\varrho\sigma} - q_{\sigma\varrho}) (g^{\mu\varrho} g^{\nu\sigma} - g^{\mu\sigma} g^{\nu\varrho}).
\end{equation}

\noindent In this case $\alpha$ is a very small number, according to Einstein's conception, equal to the gravitational constant multiplied by $\frac{8\pi}{c^2}$, that is, in the units customary for physicists, a very small number:
$$ - \alpha = 1,87 \cdot 10^{-27}; $$
I express this numerical value explicitly; this is to show that Maxwell's theory of the electron-free space, which sets $\alpha = 0$ and does not speak of $K$ at all, can be regarded as a sufficient approximation to the new approaches discussed here for ordinary measurements. See further below, No.~5.

2. I now first form purely formally the variations of the integrals $I_1$, $I_2$, which correspond to an arbitrary modification of the $g^{\mu\nu}$, $q_\varrho$ by%
\footnote{We make the assumption that $\delta g^{\mu\nu}$, $\delta g^{\mu\nu}_\varrho$, and $\delta q_\varrho$ vanish at the boundary of the region of integration.}
$\delta g^{\mu\nu}$, $\delta q_\varrho$, and write them abbreviated as follows:
\begin{subequations}%4ab
\begin{eqnarray}
\delta I_1 &=& \int \sum_{\mu,\nu} K_{\mu\nu} \delta g^{\mu\nu} d\omega
\\
\delta I_2 &=& \alpha \int \Bigl(\sum_{\mu,\nu} Q_{\mu\nu} \delta g^{\mu\nu} + \sum_\varrho Q_\varrho \delta q^\varrho\Bigr) d\omega.
\end{eqnarray}
\end{subequations}
Here $K_{\mu\nu}$, $Q_{\mu\nu}$ denote the well-known tensors, which are contragredient to the products $dw^\mu dw^\nu$:
\begin{subequations}%5ab
\begin{eqnarray}
K_{\mu\nu} &=& \left( \frac{\partial{\sqrt{g} K}}{\partial g^{\mu\nu}} - \sum_\varrho \frac{\partial \left( \frac{\partial{\sqrt{g} K}}{\partial g^{\mu\nu}_\varrho} \right)}{\partial w^\varrho} + \sum_{\varrho,\sigma} \frac{\partial^2 \left( \frac{\partial{\sqrt{g} K}}{\partial g^{\mu\nu}_{\varrho\sigma}} \right)}{\partial w^\varrho \partial w^\sigma} \right):\sqrt{g},
\\
Q_{\mu\nu} &=& \left( \frac{\partial \sqrt{g} Q}{\partial g^{\mu\nu}} \right):\sqrt{g},
\end{eqnarray}
\end{subequations}
however $Q^\varrho$ is the vector cogredient to  $dw^\varrho$:
\begin{equation}%6
Q^\varrho = - \left( \sum_\sigma \frac{\partial \left( \frac{\partial \sqrt{g} Q}{\partial q_{\varrho\sigma}} \right)}{\partial w^\sigma} \right):\sqrt{g}.
\end{equation}
\textit{The equations
\begin{equation}%7
Q^\varrho = 0,
\end{equation}
written in the coordinates $w$, are the Maxwell equations corresponding to our physical presuppositions; on the other hand the $Q_{\mu\nu}$, as you mentioned on page 407 of your note, are the energy components of the electromagnetic field.}
\newpage
%-------------------------------------------------- --------

XXXI. On Hilbert's first note on the Foundations of Physics. 555
\bigskip

3. For the sake of clarity, in advance I will distinguish between the scalar divergence of a ``vector $p^\varrho$'' and the vectorial divergence of a ``tensor $t_{\mu\nu}$.'' In our general coordinates $w^\nu$, the former is known to be expressed by the sum of:
\begin{equation}
\sum_\nu \frac{\partial (\sqrt{g} p^\nu)}{\partial w^\nu}:\sqrt{g}, %(8)
\end{equation}
but the latter is somewhat more complicated: its four components are:
\begin{equation}  %(9)
\left( \sum_{\mu,\nu} \frac{\partial(\sqrt{g} t_{\mu\sigma} g^{\mu\nu} )}{\partial w^\nu} + \frac12 \sqrt{g} \sum_{\mu,\nu} t_{\mu\nu} g^{\mu\nu}_\sigma \right):\sqrt{g}
\end{equation}
for $\sigma = 1, 2, 3, 4$.

4. I now develop \textit{the four simple partial differential equations} which $I_1$ and $I_2$ respectively satisfy
(because both are invariants under arbitrary
transformations of $w$). For this purpose, as in particular by Lie in his numerous relevant publications, the formal changes which can be made in any infinitesimal transformation are
\begin{equation}%10
\delta w^I = p^I, \dots, \delta w^{IV} = p^{IV} %(10)
\end{equation}
($p^\sigma$ is an infinitesimal vector whose higher powers may be neglected). --- You have done this for the integral $I_1$ on pp 398--400 of your note in such a way that you consider the comparatively complicated changes of $K$ in order to ascend from there through integration to the change of $I_1$.
The whole simplification of the idea is that I shall follow the formula (4a), that is, calculate the change of $I_1$ directly from the Lagrange derivative. \textit{The change of $I_1$ has to be zero, if I  (in 4a) use for the $\delta g^{\mu\nu}$ those values which correspond to the infinitesimal transformation (10)}. Since the $g^{\mu\nu}$ are cogredient with $dw^\mu dw^\nu$, one simply finds
\begin{equation}%(11)
\delta g^{\mu\nu} = \sum_\sigma (g^{\mu\nu}_\sigma p^\sigma - g^{\mu\sigma} p^\nu_\sigma - g^{\nu\sigma} p^\mu_\sigma).
\footnote{[This is explained in more detail in \S 1 of the following chapter XXXII.]}
\end{equation}
Therefore we have [by setting the $p^\sigma$ at the boundary equal to zero]:
$$ \int \sum_{\mu,\nu} K_{\mu\nu} \left( \sum_\sigma g^{\mu\nu}_\sigma p^\sigma - \sum_\sigma g^{\mu\sigma} p^\nu_\sigma - \sum_\sigma g^{\nu\sigma} p^\mu_\sigma \right) d\omega = 0. $$
\newpage

%-------------------------------------------------- -------------------------------------

556 On the Erlangen Program.
\bigskip

Here, we transform the terms $p^\mu_\sigma$, $p^\nu_\sigma$ in the known manner by partial integration, while subjecting the otherwise arbitrary $p^\sigma$ to the condition of having first and second differential quotients vanishing at the boundaries of the integration. We then get
$$ \int \sum_\sigma p^\sigma \left( \sqrt{g} \sum_{\mu,\nu} K_{\mu\nu} g^{\mu\nu}_\sigma + 2 \sum_{\mu,\nu} \frac{\partial (\sqrt{g} K_{\mu\sigma} g^{\mu\nu})}{\partial w^\nu} \right) dw^I \dots dw^{IV} = 0, $$
and from this, because of the arbitrariness of the $p^\sigma$, \textit{the four differential equations for the tensor $K_{\mu\nu}$, established by you (and Einstein):}
\begin{equation}%(12)
\sqrt{g} \sum_{\mu,\nu} K_{\mu\nu} g^{\mu\nu}_\sigma + 2 \sum_{\mu,\nu} \frac{\partial (\sqrt{g} K_{\mu\sigma} g^{\mu\nu})}{\partial w^\nu} = 0   \quad (\sigma = 1, 2, 3, 4)
\end{equation}
which we can clearly interpret as saying: \textit{the vectorial divergence of the tensor $K_{\mu\nu}$ vanishes}.

The integral $I_2$ will be treated in exactly the same way. In addition to the increments (11) of the $g^{\mu\nu}$, then only the following increments of the $q_\varrho$ occur:%
\footnote{My $\delta g^{\mu\nu}$ (11) and $\delta q_\varrho$ (13) are nothing other than that designated, in p.~398 of your note, by $p^{\mu\nu}$ and $p_\varrho$, respectively.}
\begin{equation}
\delta q_\varrho = \sum_\sigma (q_{\varrho\sigma} p^\sigma + q_\sigma p^\sigma_\varrho).
\end{equation}
\textit{We thus obtain the following four differential equations for the} $Q_{\mu\nu}$, $Q^\varrho$:
\begin{equation}%(14)
\sum_{\mu,\nu} \left( \sqrt{g} Q_{\mu\nu} g^{\mu\nu}_\sigma + 2 \frac{\partial (\sqrt{g} Q_{\mu\sigma} g^{\mu\nu})}{\partial w^\nu} \right) + \sum_\varrho \left( \sqrt{g} Q^\varrho g_{\varrho\sigma} - \frac{\partial (\sqrt{g} Q^\varrho q_\sigma)}{\partial w^\varrho} \right) = 0
\end{equation}
\qquad for $\sigma = 1, 2, 3, 4$.

It is unnecessary to put them into words. But it is well worth considering a transformation which they allow because of the special form of our $Q$ (and mutatis mutandis also occurs at different points in your note). $Q$ depends only on the differences $q_{\varrho\sigma} - q_{\sigma\varrho}$ and therefore, as a glance at (6) shows, one has a vanishing scalar divergence:
$$ \sum_\varrho \frac{\partial (\sqrt{g} Q^\varrho)}{\partial w^\varrho} = 0. $$

\textit{As a result, we can put the differential equations (14) into another form:}
$$ (14') \qquad \qquad
\sum_{\mu,\nu} \left( \sqrt{g} Q_{\mu\nu} g^{\mu\nu}_\sigma + 2 \frac{\partial (\sqrt{g} Q_{\mu\sigma} g^{\mu\nu})}{\partial w^\nu} \right) + \sum_\varrho \left( \sqrt{g} Q^\varrho (q_{\varrho\sigma} - q_{\sigma\varrho}) \right) = 0 $$
\qquad \qquad for $\sigma = 1, 2, 3, 4$.
\newpage
%-------------------------------------------------- -----------------------

XXXI. On Hilbert's first note on the Foundations of Physics. 557

\bigskip

5. From now on I introduce the basic assumption of Einstein's theory, most preferably, in the form chosen by you in your note, which is said to be \textit{that the variation should be
\begin{equation}%(15)
\delta I_1 + \delta I_2 = 0
\end{equation}
for  arbitrary $\delta g^{\mu\nu}$, $\delta q_\varrho$}.

According to (4a), (4b), this gives the known 14 ``\textit{field equations}'', namely the ten equations:
\begin{subequations}%(16a)
\begin{equation}
K_{\mu\nu} + \alpha Q_{\mu\nu} = 0
\end{equation}
and the four equations
\begin{equation} %(16b)
Q^\varrho = 0.
\end{equation}
\end{subequations}

You remark in your note, that there must be four dependencies between these fourteen equations, and show on p.~406, through special calculations, the relationship between the four equations (16b) --- the Maxwell equations --- and the ten equations (16a). Naturally, this has been already included by me in the formulas of the previous numbers. In fact, it is only necessary to add the equations (14') multiplied by $\alpha$ to the equations (12), in order to immediately deduce the vanishing of the $Q^\varrho$ as a consequence of the equations (16a).

At the same time, it is clear what was said about the character of the old Maxwellian theory as a limit of the new theory. If we treat the old Maxwellian theory in arbitrary curvilinear coordinates $w^I \dots w^{IV}$, we have always to do with a $ds^2$ whose Riemannian curvature vanishes identically, for which, therefore, the $K_{\mu\nu}$ simply are zero. On the other hand, $\alpha = 0$ is taken. \textit{Thus the ten equations (16a) are fulfilled by themselves; the energy components $Q_{\mu\nu}$ of the electromagnetic field are no longer subject to any restriction}. There remains only the four equations (16b), i.e. the Maxwell equations. As a consequence of these, the $Q_{\mu\nu}$ according to (14) have a vanishing vectorial divergence.

Of course, before Einstein, we have introduced curvilinear coordinates $w$ in physics only in such a way that we have arbitrarily transformed the three spatial coordinates, but left $t$ essentially unchanged. To include the $t$ equally in the coordinate transformation,
appears as one great achievement of Einstein.  Another is then, of course, that gravitation can be taken into account by substituting for the $ds^2$ with vanishing Riemann curvature a more general $ds^2$. --- On the other hand, in order to emphasize this, the mathematical armament for the processing of
\newpage

558 On the Erlangen Program.
\bigskip

\noindent these new physical ideas was long ago made available,
since spaces of any number of dimensions with arbitrary arc elements have been familiar to us since Riemann. This is not the place for a historical excursion that would begin with the methods of Lagrange's Mecanique analytique, where in addition to the ever-recurring work of Christoffel, those of Beltrami and Lipschitz should be discussed.

6. Now, without using the field equations (16), I want to add the equations (12), (14) together after multiplying the latter by $\alpha$. This gives for $\sigma = 1, 2, 3, 4$ the identities:
\begin{eqnarray}%(17)
\sum_{\mu,\nu} \sqrt{g} (K_{\mu\nu} + \alpha Q_{\mu\nu}) g^{\mu\nu}_\sigma + \sum_\varrho \alpha \sqrt{g} Q^\varrho q_{\varrho\sigma}
\\ \nonumber
= -2 \sum_{\mu,\nu} \frac{\partial [\sqrt{g} (K_{\mu\sigma} + \alpha Q_{\mu\sigma}) g^{\mu\nu} - \frac{\alpha}2 Q^\nu q_\sigma]}{\partial w^\nu}.
\end{eqnarray}
I multiply these equations by $p^\sigma$ (where by $p^\sigma$, an arbitrary vector cogredient to $dw^\sigma$ is understood), and sum according to $\sigma$. Here, on the right hand, I can take the $p^\sigma$ inside the differentiation signs by placing the corresponding supplementary terms on the left hand. On the other hand, I exchange the index names $\sigma$ and $\nu$ on the left hand, and instead of $2 g^{\mu\sigma} p^\nu_\sigma$, use the symmetrical $g^{\mu\sigma} p^\nu_\sigma + g^{\nu\sigma} p^\mu_\sigma$, which is equivalent in the context of these considerations. Thus, the following is produced:
\begin{eqnarray}%(18)
\sum_{\mu,\nu,\sigma} \sqrt{g} (K_{\mu\nu} + \alpha Q_{\mu\nu}) (g^{\mu\nu}_\sigma p^\sigma - g^{\mu\sigma} p^\nu_\sigma - g^{\nu\sigma} p^\mu_\sigma)
\\
+ \sum_{\varrho,\sigma} \alpha \sqrt{g} Q^\varrho (q_{\varrho\sigma} p^\sigma + q_\sigma p^\sigma_\varrho)
\nonumber\\
= -2 \sum_{\mu,\nu,\sigma} \frac{\partial \{
   [ \sqrt{g} (K_{\mu\sigma} + \alpha Q_{\mu\sigma}) g^{\mu\nu} - \frac\alpha2 \sqrt{g} Q^\nu q_\sigma ] p^\sigma \}}
   {\partial w^\nu}, \nonumber
\end{eqnarray}
which naturally is only another way of writing (17). In view of the particular value which I have assumed for your $H$ from the beginning ($H = K + \alpha Q$), the left hand here is exactly what you give as the value of the scalar divergence of your \textit{energy vector} $e^\nu$ multiplied by $\sqrt{g}$ (p 402 in your Note), thus
$$ \sum_\nu \frac{\partial \sqrt{g} e^\nu}{\partial w^\nu}. $$
It follows that your \textit{energy vector $e^\nu$ from}
$$ \Bigl( -2 \sum_{\mu,\sigma} (K_{\mu\sigma} + \alpha Q_{\mu\sigma}) g^{\mu\nu} + \frac\alpha2 \sqrt{g} Q^\nu q_\sigma \Bigr) p^\sigma $$
\bigskip
[Editor's note: A missing left parenthesis has been inserted.]

\newpage

XXXI. On Hilbert's first note on the Foundations of Physics. 559
\bigskip

\noindent\textit{only differs by a term whose scalar divergence vanishes identically}.

If we now take the 14 field equations (16a), (16b), then $e^\nu$ is reduced to this additional term and the statement on p.~402 of your note, that
\begin{equation}%(19)
\sum_\nu \frac{\partial \sqrt{g} e^\nu}{\partial w^\nu} = 0
\end{equation}
holds, appears as an identical statement. This statement cannot, therefore, be regarded as an analogy to the conservation law of energy, as is the case in ordinary mechanics. For if we write in the latter:
$$ \frac{d (T + U)}{dt} = 0, $$
then this differential relation is not identically satisfied, but only as a result of the differential equations of mechanics.

7. Of course, it would be desirable to specify explicitly the additional terms on which your $e^\nu$ differs from the vanishing elements due to the field equations. But I find your formulas so complicated that I have not done the calculations. Only this seems clear: that it comprises components linear in the $p^\sigma$, others containing the $p^\sigma_\mu$, and perhaps those which contain the $p^\sigma_{\mu\nu}$ linearly. It should not be difficult to specify the most general vectors of this type whose scalar divergence vanishes identically. We obtain generally vectors $X^\nu$ of vanishing divergence by starting from any six-tensor
(an obliquely symmetrical tensor) $\lambda^{\mu\nu}$ and setting
\begin{equation}%(20)
\sqrt{g} X^\nu = \sum_\mu \frac{\partial \lambda^{\mu\nu}}{\partial w^\mu}.
\end{equation}
If one wants to have linearity of the $X^\nu$ in the $p^\sigma$ and the $p^\sigma_\mu$, then one can for example choose
\begin{equation}%(21)
\lambda^{\mu\nu} = \left( \bigl( \sum g^{\mu\varrho} q_\varrho \bigr) p^\nu - \bigl( \sum g^{\nu\varrho} q_\varrho \bigr) p^\mu \right).
\end{equation}

8. Here I have to make a substantial digression. You know that Miss~N\"other continues to advise me on my work, and that I have only penetrated into the present matter through her. When I was  speaking recently to Miss N\"other of my results concerning your energy vector, she was able to inform me that she had derived the same results on the basis of your note (and thus not from the simplified calculations of my number 4) more than a year ago, and put it in a manuscript (which I was later able to read); only she had not stated it as forcefully as I had done recently at the Mathematical Society (22. January).

\newpage

560 On the Erlangen Program.
\bigskip

9. Finally, I would like to draw your attention to the fact that for the ``conservation laws'' as Einstein formulated in 1916,\footnote{Compare the independently published work: The Foundations of the General Theory of Relativity (Leipzig, 1916) and, in particular, the communication to the Berlin Academy of 29 October 1916, ``Hamilton's Principle and General Relativity Theory'' (Session Reports, pp. 1111--1116).}
the same is true for your proposition (19). He is actually stating it completely clearly. I will not go into the details of his calculation here, but will only refer to his conclusion, which he writes as follows:
\begin{equation}%(22)
\sum_\nu \frac{\partial}{\partial w^\nu} (\mathfrak{T}^\nu_\sigma + \mathfrak{t}^\nu_\sigma) = 0, \qquad (\sigma = 1, 2, 3, 4),
\end{equation}
where the $\mathfrak{T}^\nu_\sigma$ and the $\mathfrak{t}^\nu_\sigma$ are referred to as the ``mixed'' energy components of the electromagnetic or gravitational field. He states that these $\mathfrak{T}^\nu_\sigma+\mathfrak{t}^\nu_\sigma$ can be expressed using the field equation as follows by means of a function $\mathfrak{G}^*$ dependent on the coordinate system:
\begin{equation}%(23)
\mathfrak{T}^\nu_\sigma + \mathfrak{t}^\nu_\sigma = -\sum_{\mu,\varrho} \left( \frac{\partial}{\partial w^\varrho} \left( \frac{\partial \mathfrak{G}^*}{\partial g^{\mu\sigma}_\varrho} g^{\mu\nu} \right) \right),
\end{equation}
and that for this $\mathfrak{G}^*$ independently of the value of $\sigma$ there is the identical equation:
\begin{equation}%(24)
\sum_{\mu,\nu,\varrho} \frac{\partial^2}{\partial w^\nu \partial w^\varrho} \left( \frac{\partial \mathfrak{G}^*}{\partial g^{\mu\sigma}_\varrho} g^{\mu\nu} \right) = 0.
\end{equation}
That is exactly what matters.

In order to establish the connection with the terms used by me, I note that Einstein's $\mathfrak{T}^\nu_\sigma$ are the same as my $\sum_\mu \sqrt{g} Q_{\mu\sigma} g^{\mu\nu}$, Einstein's $\mathfrak{t}^\nu_\sigma$ may deviate from the corresponding
$\frac1\alpha \sum_\mu \sqrt{g} K_{\mu\sigma} g^{\mu\nu}$ by a summand which results if the equations (23) are compared with the field equations
$$ K_{\mu\nu} + \alpha Q_{\mu\nu} = 0. $$

\bigskip

\hfil \textbf{II. From the answer of D. Hilbert.} \hfil

\dots I agree with your comments on the energy law: Emmy N\"other, upon whose help I called on for more than a year to clarify such analytical questions concerning my energy theorems, found at that time that the energy components I had set up---as well as those of Einstein---can be converted, formally using the Lagrange differential equations (4), (5) of my first note, into expressions whose divergence is \textit{identical}, that is without using Lagrange's equations (4), (5) it vanishes.

\newpage
%-------------------------------------------------- --------------------------

XXXI. On Hilbert's first note on the Foundations of Physics. 561

\bigskip

\setcounter{equation}{0}

\noindent On the other hand, since the energy equations of classical mechanics, elasticity theory, and electrodynamics are only satisfied with the Lagrange differential equations of the problem, then it is justified if you do not see the analogy with these theories in my energy equations. To be sure, I maintain that for \textit{general} relativity, i.e. in the case of the \textit{general} invariance of the Hamiltonian function, energy equations which in your sense correspond to the energy equations of the orthogonally invariant theories do not exist at all; indeed, I would even call this fact a characteristic feature of the general theory of relativity. For my assertion the mathematical proof should be possible.

On this occasion, please allow me to briefly explain how, in my lecture of the last winter, I dealt with the energy equations of the orthogonal invariant theories of physics (electrodynamics, hydrodynamics, and elasticity theory).

For the sake of brevity, let us assume that the world function $H$ as an orthogonal invariant which depends only on the components of the electrodynamic four-potentials $q_s$ and their first derivatives $q_{sl}$ on $w_k$ ($s, l = 1, 2, 3, 4$), --- The methods apply in the same way, if $H$ is a four-fold density of $r$ and its derivatives, or else of other physical parameters, along with their derivatives,---: then the Hamiltonian principle is
\begin{equation}
\delta \int Hd\omega = 0.
\end{equation}
The system of the four Lagrangian differential equations is
\begin{equation}
[H]_s = 0, \qquad (s = 1, 2, 3, 4)
\end{equation}
where
$$ [H]_s = \frac{\partial H}{\partial q_s} - \sum_k \frac{\partial}{\partial w_k} \frac{\partial H}{\partial q_{sk}}. $$

In order to arrive at the energy equations of this problem, we propose the path which the statements of my first communication pointed out, namely, the path through gravitation theory. Let $\bar H$ be a general invariant with the arguments
$$ q_s, q_{sl}, g^{\mu\nu}, g^{\mu\nu}_l $$
which for
\begin{equation}
g^{\mu\nu} = g_{\mu\nu} = \delta_{\mu\nu}, \qquad g^{\mu\nu}_l = 0
\end{equation}

 ---------------------------

{\tiny{Klein, Collected math. Treatises. I. 36}}

\newpage
562 On the Erlangen Program.
\bigskip

into $H$ transforms; we obtain this by substituting the covariant derivatives
$$ \bar q_{sl} = q_{sl} - \sum_h \left\{ sl \atop h \right\} q_h $$
instead of $q_{sl}$ and at the same time performing the convolution with $g^{\mu\nu}$. For example, if $H$ contains the orthogonally invariant expression
\begin{equation}
- Q = \sum_{m,n} (q_{mn} - q_{nm})^2 = \frac14 \sum_{m,n} M_{mn}^2,
\end{equation}
then it must be replaced by
$$ - \bar Q = \frac14 \sum_{m,n,k,l} M_{mn} M_{kl} g^{mk} g^{nl}. $$
The expression
$$ T = \sum_{s,h} q_{sh}^2 $$
is to be converted into
$$ \bar T = \sum_{s,h,m,n} \bar q_{sh} \bar q_{mn} g^{sm} g^{hn}, $$
and so on.

Now, for every general invariant there is an identity, which in my first communication (Theorem III) has been proved only in the case that the invariant depends on the $g^{\mu\nu}$ and its derivatives; but the invoked proof method also applies to our general invariant $\bar H$. Using the nomenclature of my first note, we get instead of the equation there,
$$ \int P_g (J \sqrt{g}) d\omega = 0 $$
in our case the equation
$$ \int \{ P_g (\bar H \sqrt{g}) + P_q (\bar H \sqrt{g}) \} d\omega \equiv \int \{ P (\bar H \sqrt{g}) \} d\omega = 0, $$
an identity that immediately results is
$$ \int \Bigl\{ \sum_{\mu,\nu} [\sqrt{g} \bar H]_{\mu\nu} p^{\mu\nu} + \sum_\mu [\sqrt{g} \bar H]_{\mu} p_{\mu} \Bigr\} d\omega = 0. $$
After the introduction of $p^{\mu\nu}$, $p_\mu$, and the application of partial integration, we can bring the integral of the left hand side to a form in which the integrand is multiplied by $p^s$; but since $p^s$ is an arbitrary vector, the other factor under the integral sign must be identically zero, and this gives the identities ($s = 1, 2, 3, 4$):%
\footnote{[See in this case my formula (14), which agrees with this term by term. K.]}
\begin{eqnarray}%(5)
\sum_{\mu,\nu} [\sqrt{g} \bar H]_{\mu\nu} g^{\mu\nu}_s &-& 2 \sum_m \frac{\partial}{\partial w_m} \Bigl\{ \sum_\mu [\sqrt{g} \bar H]_{\mu s} g^{\mu m} \Bigr\}
\\ \nonumber
&+& \sum_\mu [\sqrt{g} \bar H]_\mu q_{\mu s} - \sum_\mu \frac{\partial}{\partial w_\mu}([\sqrt{g} \bar H]_{\mu} q_s ) = 0.
\end{eqnarray}

\newpage

XXXI. On Hilbert's first note on the Foundations of Physics. 563

\bigskip

These four identities are, as you have already pointed out above, those whose existence is asserted in my Theorem I, between the 14 Lagrangian equations of our problem.

If we now return to the original problem (1) by eliminating the gravitational potentials based on (3) and taking into account Lagrange's differential equations (2), the identities (5) go over to
\begin{equation}%(6)
\sum_m \frac{\partial}{\partial w_m} \{ [\sqrt{g} \bar H]_{ms} \}_{g_{\mu\nu} = \delta_{\mu\nu}} = 0.
\end{equation}
If we denote the bracketed terms
\begin{equation}%(7)
\varepsilon_{ms} = 2 \{ [\sqrt{g} \bar H]_{ms} \}_{g_{\mu\nu} = \delta_{\mu\nu}}
\end{equation}
as the \textit{components of the energy tensor}, we obtain the desired energy equations of the physical problem (1) in the divergence equations (6).

If we take especially for $H$ the invariant $Q$ in (4), then $\varepsilon_{ms}$ are the components of the known electromagnetic energy tensor, and because of the Maxwell equations
$$ \{ [\sqrt{g} \bar H]_m \}_{g_{\mu\nu} = \delta_{\mu\nu}} = {\rm Div}_m M = r_m $$
--- by $r$ the electric four  current density is understood --- in this case our identities (5) become
$$ {\rm Div}_s \varepsilon - \sum_m r_m q_{ms} + \sum_m \frac{\partial}{\partial w_m} (r_m q_s) = 0 $$
or because of ${\rm Div}_s \, r = 0$:
$$ {\rm Div}_s \varepsilon = - r_s \cdot M, $$
i.e., it provides the well-known divergence expression for the ponderomotive force.

Only in the case of general relativity, that is, when the original invariant $H$ is a general invariant, the given path for producing energy equations for problem (1) fails. In the general theory of relativity, as a substitute for the missing energy equations in your sense, we have the fact of the fourfold excess of the Lagrange equations (theorem I of my first communication), as expressed above in the four identities (5). Conversely, the energy statement of the orthogonally invariant theories appear as the residue of those four identities of gravitation theory.
\bigskip

\hfill 36 *
\newpage

564 On the Erlangen Program.
\bigskip

It should be noted that the energy tensor (7) not only, as is immediately evident, has the properties of orthogonal invariance and symmetry, but also, in addition, the requirements of the particular physical theory are met each time: in the case of electrodynamics, $H$ which only contains $q_{sl}$ in the compound
$$ M_{ks} = q_{sk} - q_{ks}, $$
also depends only on these components of the six-vector $M$, and, on the other hand, in the case of the elasticity theory, it also depends only on the actual distortion variables which occur in the questions of elasticity. \dots

\bigskip

\hfil \textbf{III. From another letter by F. Klein.}

\dots It is up to me to characterize the difference between the orthogonal-invariant theory of electrodynamics and the gravitational force.

In this regard, it is especially clarifying, if one, as I indicated above (no. 5), as an intermediary, turns to the treatment of classical electrodynamics in arbitrary (``curvilinear'') world coordinates.

Your main statement, that the energy components of the electrodynamic field are simply represented by the $Q_{\mu\nu}$, then comes to the fore in its entire meaning; I would therefore prefer not to rely on modern gravitation theory in this theorem.

Also, I find it useful to keep the integrals $\int Kd\omega$ and $\int Qd\omega$ apart in the representation, and not to merge from the beginning to an integral $\int Hd\omega$.

We then have four identities for the $K_{\mu\nu}$ and the $Q_{\mu\nu}$ [the equations (12) and (14) --- or (14') --- of my first letter], and hence on the whole \textit{eight}, and the contrast of the earlier and the present theory can then be formulated in precise sentences as follows:

1. In both cases in addition to the eight identities we have 14 ``field equations'' for the comparison that is considered here.

2. These are in the earlier theory
$$ {\rm a}) \quad K_{\mu\nu} = 0,\footnote{[As a result of the 20 equations which show the identical disappearance of Riemann's curvature. K.]}
\qquad {\rm b}) \quad Q^\varrho = 0. $$
By virtue of the ten equations a) and the four equations b), the four identities (12) are fulfilled by themselves, the identities (14) --- or (14$'$)---however are reduced by virtue of the four equations (b)  to the four statements which are called the four conservation statements (momentum-energy).

\newpage
%-------------------------------------------------- -------

XXXI, Hilbert's first note on the Foundations of Physics. 565

\bigskip

3. Therefore in the new theory one has the field equations
$$ {\rm a}') \quad K_{\mu\nu} + \alpha Q_{\mu\nu} = 0, \quad \mathrm{with}\; \alpha \ne 0, \qquad {\rm b}') \quad Q^\varrho = 0. $$
Now the equations $Q^{\varrho} = 0$ by virtue of the eight identities appear as a result of the ten equations (a$'$).

From the identities (14), if the $Q^\varrho$ is omitted, they are still ``conservation laws'' for the $Q_{\mu\nu}$. But these now have no independent (physical) meaning, because they reduce by virtue of the ten equations (a$'$) to the four identities (12); they are already included in the ten field equations.

All this is factually in full agreement with the expositions of your letter. However, I would be very interested to see the execution of the mathematical proof that you promised at the end of the first paragraph of your answer. .......

\bigskip
----------------------
\bigskip

[The stated demonstration has now been provided by Miss E. N\"other, see her note on ``Invariant Variational Problems'' in the G\"ottingen news of 26 July 1918. I return to this at the close of XXXII.

Moreover, in order to make clear the relations between the articles XXXI to XXXIII on the Erlangen program, I would like to add the following remarks:

1. The invariant theory of the Lorentz group treated in XXX is precisely what modern physicists call the ``special theory of relativity''.

2. The Lorentz group can be seen to be the largest continuous family of the most general continuous transformations for the finite values of the $x, y, z, t$, which transform the quadratic differential form
$$ ds^2 = dt^2 - \frac{dx^2 + dy^2 + dz^2}{c^2} $$
into itself.

3. Let us suppose, instead of the $xy zt$, that some real, uniformly differentiable, finitely continuous functions
$$ w^\varrho = \varphi^\varrho (x, y, z, t) \qquad\qquad (\varrho = 1, 2, 3, 4) $$
are introduced.
In this way, the $ds^2$, which has just been written, should pass into a more general quadratic form of the $dw$, which we shall write immediately in Einstein's way:
$$ ds^2 = \sum g_{\mu\nu} dw^\mu dw^\nu. $$

4. This new $ds^2$ is, of course, the same as that given under 2. with the inertial character $+ - - -$. Its coefficients $g_{\mu\nu}$ are continuous, sufficiently differentiable, real functions of the $w$, which are only particularized in that the Riemannian curvature formed for $ds^2$ vanishes identically.

\newpage

566 On the Erlangen Program.

\bigskip

5. According to the principles of the Erlangen program, we can now also treat the special theory of relativity in the way that we based the entire group of all real, continuous, sufficiently often differentiable, unambiguously reversible transformations of the $w^\varrho$, but \textit{adjoining} the $ds^2$ of 3., i.e., to accept the changes which the $g_{\mu\nu}$, in the respective transformations of the $w$, suffer. One obtains unambiguously determined linear transformations of the $g_{\mu\nu}$, since the relations to which the $g_{\mu\nu}$ are bound as coefficients of a form of vanishing degree of curvature are of too high a character to be influential. Moreover, it should be noted that not only the substitution coefficients, but also the $g_{\mu\nu}$, are functions of $xyzt$ and $w^\varrho$, respectively. From this we obtain the changes which the differential quotients of the $g_{\mu\nu}$ undergo in the respective transformation. The group ``expanded'' by all these formulas is to be taken as a basis.

6. If we do this, we have made a decisive step towards the ``general theory of relativity.'' A further step will be that for the coefficients $g_{\mu\nu}$ of the $ds^2$, we introduce the most general, for real $w$ everywhere realizable, sufficiently differentiable functions of $w$. Then
the Riemannian curvature, and the invariant derived from it called by Hilbert $K$, are no longer identically zero.

For the rest, the ``group'' will be selected as indicated under 5. ---

By the same token, the question arises as to the context of the world as a whole, analogous to the considerations in the case of the geometry of the plane
with respect to the considerations in
the treatise~XXI.
This question still seems to be little elaborated: certain possibilities are given in the treatise~XXXIII. In the special theory of relativity in which, in order to obtain all the points of the world, we let $xyzt$ run from $-\infty$ to $+\infty$, the whole question falls away naturally.

7. The general theory of relativity of the pure gravitational field results from Einstein's basic approach (which was formulated exactly by Einstein and Hilbert at nearly the same time%
\footnote{ Einstein ``On the General Relativity Theory" in the meeting reports of the Berlin Academy of 11 and 25 November 1915 (pages 799 to 801 respectively pages 844 to 847 of the year), Hilbert in his first comment on the ``Foundations of Physics" in the G\"ottingen News of 20 Nov. 1915 (cited above). There can be no question of a priority issue, because both authors pursue quite different ideas (in such a way that the compatibility of the results did not appear to be certain at first). Einstein proceeds inductively and immediately thinks of any material system. Hilbert, by imposing, by the way, the restriction on electrodynamics mentioned in the text, under (8), from the above-mentioned highest principles of variation. Hilbert also linked to Mie in particular. It was only in his communication to the Berlin Academy of 29 October 1916, mentioned above (p.~560), that Einstein made the connection of the two approaches.}) by comparing the $g_{\mu\nu}$ to the ten, inversely invariant equations $K_{\mu\nu} = 0$, in their totality of the group in question (For the sake of brevity, I use the designation (5a) of my own note).

8. Let us now take into consideration, aside from gravity, any further physical phenomena; or, rather, as in the preceding text, following Hilbert's first note, we limit ourselves to the electromagnetic processes in empty space.

9. We shall consider this most easily, even in the case of the special theory of relativity (which, unfortunately, is not expressed in chapter XXX), if, besides our

\newpage

XXXI. Hilbert's first note on the Foundations of Physics. 567

\bigskip
\noindent $ds^2$, we use the linear form
$$ \sum q_\varrho dw^\varrho $$
where the real, everywhere continuous, sufficiently differentiable functions $q_\varrho$, are the so-called four-potential of the electromagnetic field.

10. The group to be defined now extends to 5.\ by the fact that, besides the transformations of the $g_{\mu\nu}$ and their differential quotients induced by the transformations of the $w$, one now also has  the $q_\varrho$ and their differential quotients.

11. The $g_{\mu\nu}$, $q_\varrho$, however now are 14, and Eqs.~(16a), (16b) of the text:
$$ K_{\mu\nu} + \alpha Q_{\mu\nu} = 0, \qquad Q^\varrho = 0 $$
are invariant subjected to the extended group.
This is in conjunction with (10) the kernel of the general theory of relativity of physics, as far as this is concerned. ---

These formulations, of course, only express in other languages, what was already said in Einstein and Hilbert. I would particularly like to refer to Hilbert's second communication on the Foundations of Physics (in the G\"ottinger Nachrichten, 1917, pp.~53--76\footnote{Provided on December 23, 1916.}).
Here, on p.~61, it is expressly stated that only those conclusions which follow from the differential equations of 11.\
have a physical sense, which, like the differential equations themselves
(NB in contrast to the group defined under 10.) have an invariant character.
This is mutatis mutandis exactly what is required in the Erlangen program by the statements of any geometry (arbitrarily characterizable by a group).

It is scarcely necessary to say that the further development of Einstein's theory, as given by Weyl, can also be connected with the scheme of the Erlangen program.

There is even a particularly close relation to individual versions there (Note VI, Vol. XXVII, pp.~491--492), insofar as not a form $ds^2$, but an equation $ds^2 = 0$ is used as the basis. K.]

%---------------------------------------------

%9) Provided on December 23, 1916.

\end{document}